\documentclass[%
 aip,
 amsmath,amssymb,
 reprint,%
]{revtex4-1}
 
\usepackage{graphicx}% Include figure files
\usepackage{dcolumn}% Align table columns on decimal point
\usepackage{bm}% bold math
\usepackage[utf8]{inputenc}
\usepackage[T1]{fontenc}
\usepackage{mathptmx}
\usepackage{physics}

\begin{document}
 
%\preprint{APS/123-QED}
 
%\title{High-Q 100 GHz Superconducting Resonator for single Mm to Optical Photon conversion}% Force line breaks with \\
\title{A tunable High-Q millimeter wave cavity for hybrid circuit and cavity QED experiments}% Force line breaks with \\
	\author{Aziza Suleymanzade}
	\author{Alexander Anferov}
	\author{Mark Stone}
	\author{Ravi K. Naik}
	\author{Jonathan Simon}
	\author{David Schuster}
\affiliation{Department of Physics and James Franck Institute, University of Chicago, Chicago, IL}
\date{\today}% It is always \today, today,
\begin{abstract}

%ANCHOR FOR GENERAL FEEDBACK

The millimeter wave (mm-wave) frequency band provides exciting prospects for quantum science and devices, since many high-fidelity quantum emitters, including Rydberg atoms, molecules and silicon vacancies, exhibit resonances near 100 GHz. High-Q resonators at these frequencies would give access to strong interactions between emitters and single photons, leading to rich and unexplored quantum phenomena at temperatures above 1K. We report a 3D mm-wave cavity with a measured single-photon internal quality factor of $3  \times  10^{7}$ and mode volume of $0.14 \times   \lambda^3$ at $98.2$ GHz, sufficient to reach strong coupling in a Rydberg cavity QED system. An in-situ piezo tunability of $18$ MHz facilitates coupling to specific atomic transitions. Our unique, seamless and optically accessible resonator design is enabled by the realization that intersections of 3D waveguides support tightly confined bound states below the waveguide cutoff frequency. Harnessing the features of our cavity design, we realize a hybrid mm-wave and optical cavity, designed for interconversion and entanglement of mm-wave and optical photons using Rydberg atoms.

\end{abstract}
\maketitle

Cavity and circuit Quantum ElectroDynamics (QED) systems provide unprecedented control over photonic quantum states via coupling to strongly nonlinear single emitters. This effort began with pioneering works in Rydberg cavity QED, demonstrating first nonclassical micromaser radiation~\cite{rempe_observation_1990}, Schrodinger cat states and early EPR experiments~\cite{brune_quantum_1996, raimond_colloquium:_2001}. Since then, cavity and circuit QED systems have become essential tools for exploring quantum phenomena both in the optical~\cite{birnbaum_photon_2005, boca_observation_2004, thompson_observation_1992, imamoglu_quantum_1999} and microwave~\cite{wallraff_strong_2004, paik_observation_2011} regimes. Hybrid systems, which cross-couple these regimes, can harness the strengths of optical systems for communication and microwave systems for quantum information processing, yielding a more powerful toolset for quantum information technology~\cite{ladd_quantum_2010}. In particular, the coherent interconversion of microwave and optical photons would enable large quantum networks and robust transfer of quantum information~\cite{andrews_bidirectional_2014, hill_coherent_2012,bochmann_nanomechanical_2013,forsch_microwave--optics_2018,vainsencher_bi-directional_2016,abdo_full_2013,hafezi_atomic_2012, kiffner2016two}.

\begin{figure}[t!]
\centering
  \includegraphics[width=1\linewidth]{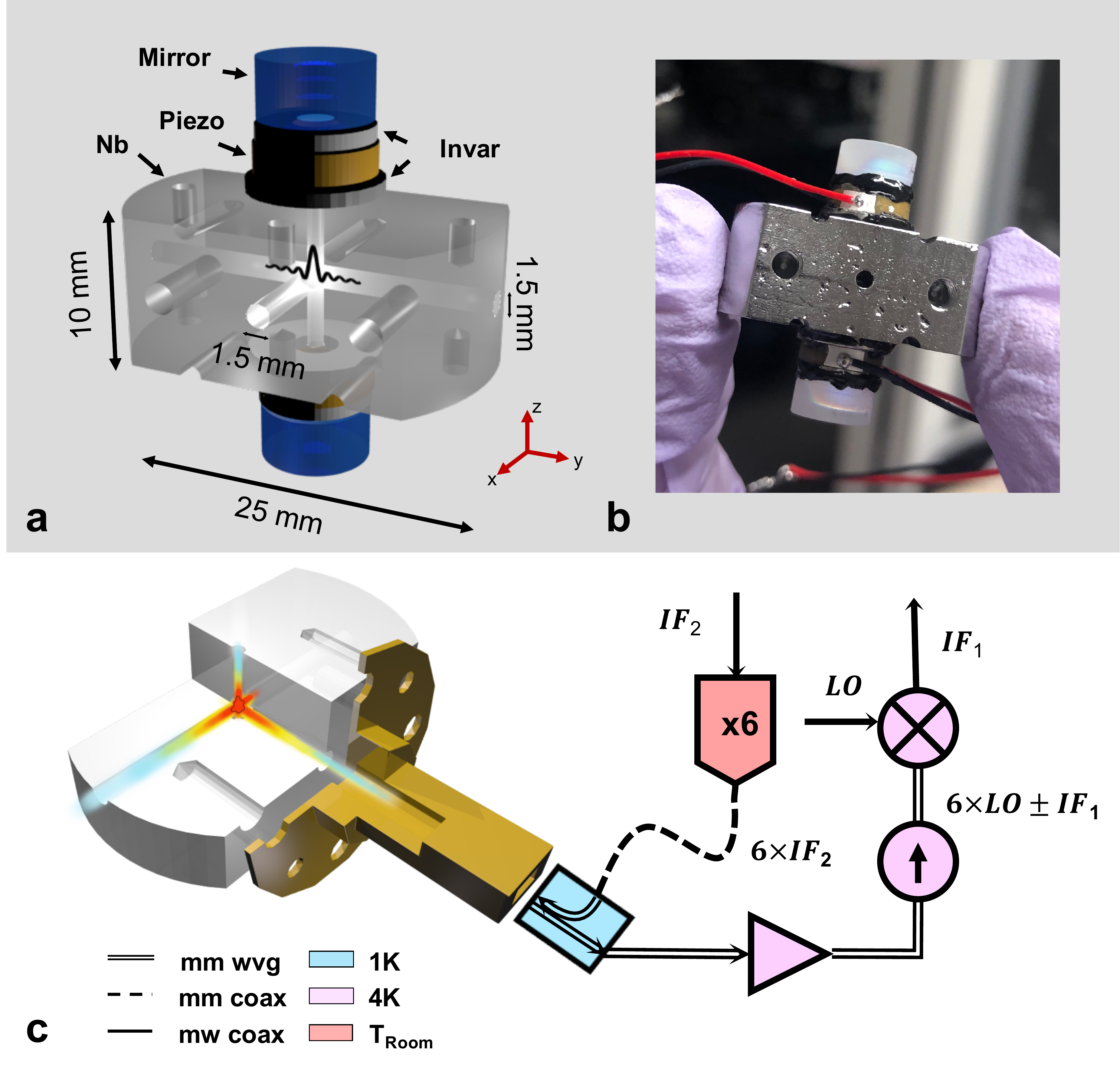}
\caption{\textbf{a.} Schematic of crossed mm-wave and optical cavities. The superconducting mm-wave cavity is formed by the intersection of three evanescent waveguides. The x-axis waveguide will be employed for atom transport, y-axis waveguide for mm-wave coupling and z-axis waveguide for an optical Fabry-P\'erot cavity. Each side of the Fabry-P\'erot cavity includes Invar spacers to prevent differential thermal contractions and a piezo actuator for tuning and locking the frequency of the optical cavity. \textbf{b.} Photograph of the assembled crossed mm-wave and optical cavity with wired piezos. \textbf{c.} Schematic reflection measurement setup for the mm-wave cavity. 100 GHz photons are delivered into the coupling port of the cavity through a WR10 waveguide.}
\label{fig:Fig1}
\end{figure}

Mm-wave frequencies provide a promising platform for hybrid quantum science~\cite{pechal_millimeter-wave_2017} at less explored, but potentially beneficial length and energy scales. Firstly, 100 GHz resonances with long coherence times are abundant among commonly used optical and microwave quantum emitters such as Rydberg atoms~\cite{li_millimeter-wave_2003}, molecules~\cite{zhou_direct_2015} and silicon vacancies~\cite{sukachev_silicon-vacancy_2017}, though they are rarely harnessed for quantum science due to lack of both high-Q resonators with tight mode confinement and mature mm-wave manipulation technology. Secondly, the mean thermal photon occupation of a 100 GHz resonator at 1K is $n_{ph} = 1/ (e^{\frac{h\nu}{k_{b}T}}-1) = 0.009 \ll 1$. This puts such a resonator in the quantum regime at temperatures accessible with simple pumped $^4$He with much larger cooling powers and lower cost and complexity than the dilution refrigerators required to reach $\sim 20$mK for $10$ GHz experiments. Finally, the intermediate length scale of mm-waves enables development of scalable high Q-factor devices using both near and far field wave engineering techniques.

More broadly, the mm-wave band is gaining interest across many fields of science and technology. Advances in mm-wave detection are essential for observational cosmology and study of the cosmic microwave background~\cite{carlstrom_10_2011,vieira_extragalactic_2010}. Terahertz and near-terahertz radiation exhibits vast potential in hazardous chemical sensing and effective medical diagnostics~\cite{ jepsen_terahertz_2011, sizov_terahertz_2018}. Recently, the search for higher bandwidth and lower latency communication brought mm-wave wave frequencies into the focus of the telecommunication industry~\cite{rappaport_millimeter_2013, roh_millimeter-wave_2014}. Once prohibitively limited and expensive, mm-wave technology is becoming more available due to this progress, encouraging the development of mm-wave quantum devices.

In this work, we develop technology to hybridize mm-wave and optical photons for quantum science applications. In particular, we report a crossed optical and mm-wave cavity, with Rydberg atoms envisioned as a transducer. Our central breakthrough is the optically open, tunable 3D mm-wave cavity with $Q =3\times 10^7 $ and mode volume $V = 0.14\lambda^3$ shown in Fig.~\ref{fig:Fig1}a.

\begin{figure}[t!]
\centering
  \includegraphics[width=1\linewidth]{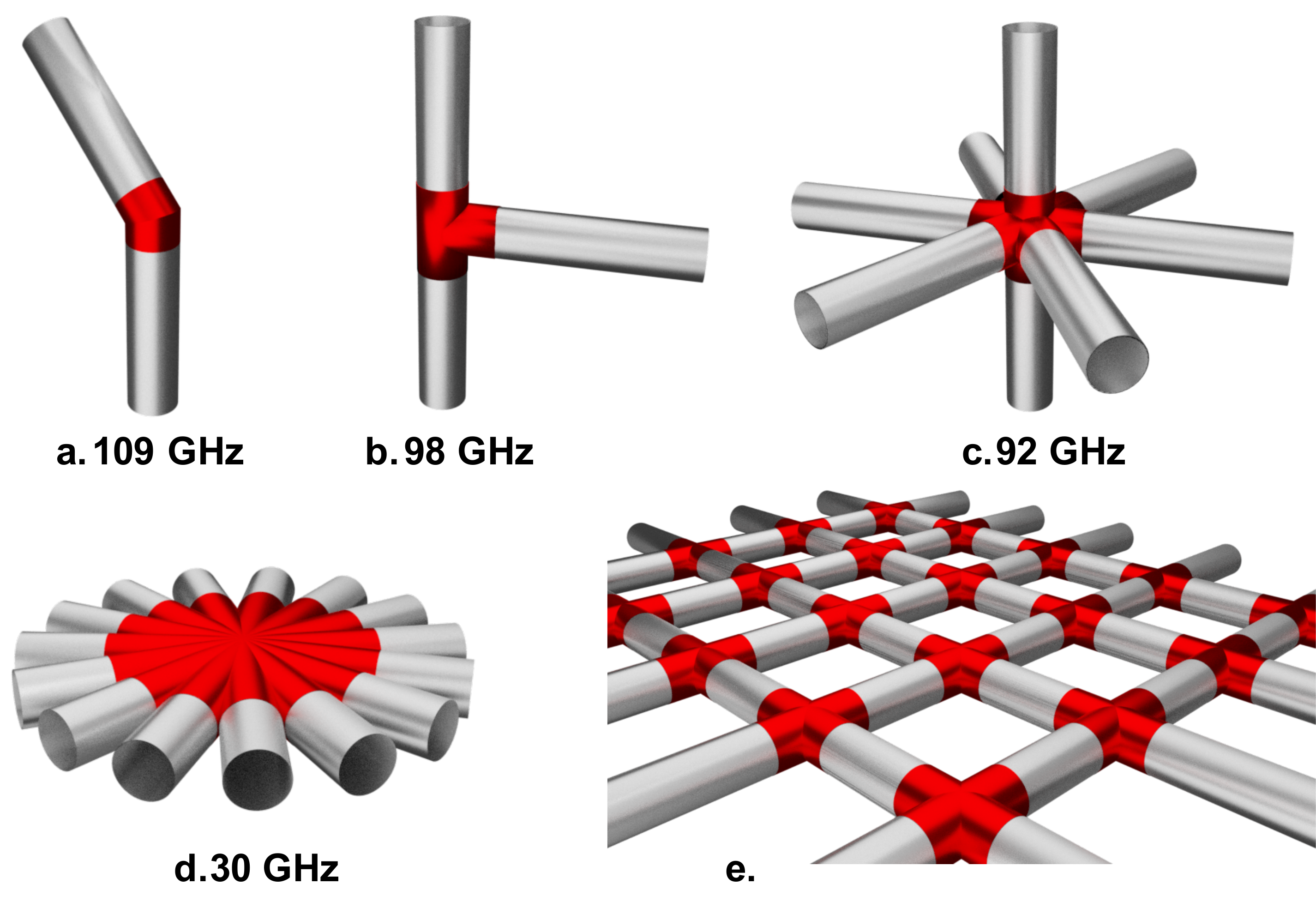}
\caption{Various seamless resonators made from intersecting 1.6mm diameter tubes. The diameter of these tubes is chosen such that the low-frequency cutoff of each tube is $\sim 110$ GHz. Examples of geometries with bound defect modes include: \textbf{a.} an ``elbow'' resonator at 109GHz made by intersection 2 tubes. \textbf{b.} a 98GHz ``tee''. \textbf{c.} 4-beam  92GHz ``star'', \textbf{d.} a 30GHz quasi-cylindrical resonator, made of 15 tubes, \textbf{e.} a 2D lattice of mm-wave ``cross'' resonators.}
\label{fig:Fig2}
\end{figure}

\begin{figure*}[t!]
  \includegraphics[width=\textwidth]{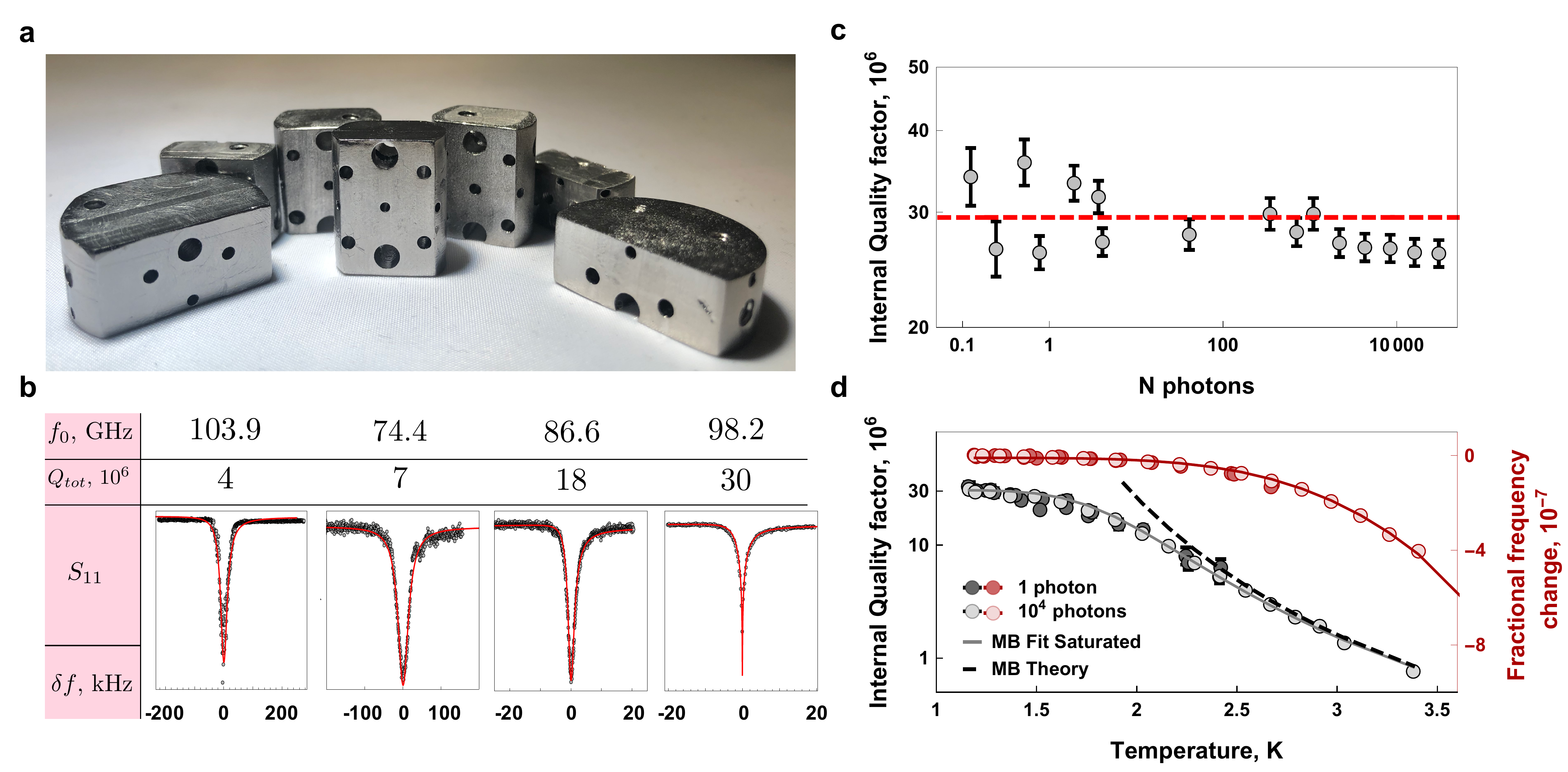}
  \caption{\textbf{a.} Photograph of various tested mm-wave cavity geometries. \textbf{b.} Reflection spectra from several cavities with varying frequencies and coupling Q's, resulting in different total Q's. \textbf{c.} Internal Q as a function of number of photons for the hybrid cavity. The constant trend indicates that the limiting loss mechanism is not power-dependent. \textbf{d.} Internal Q (black) and fractional frequency change (red) as a function of temperature for the hybrid cavity. The deviation from Mattis-Bardeen curve at 2.3K suggests that the resistivity of Nb does not limit the lifetime of the photons at the lowest temperatures.
  }
  \label{fig:Fig3}
\end{figure*}

In what follows, we first describe the design and manufacture of the seamless mm-wave cavity. We then introduce the mm-wave measurement setup in a 1K cryostat and characterize the properties of the device as a function of intra-cavity power and temperature. We also demonstrate in-situ mechanical frequency control in the cryogenic environment. Finally, we discuss the application of this device in our Rydberg hybrid experiment and more broadly, in cavity and circuit QED systems.

The key features of our device are its completely seamless design, sub-wavelength mode volume and abundant optical access to the strongly confined mm-wave mode. Typically composed of two pieces, high-Q 3D cavities are vulnerable to photon leakage through the seam between the pieces~\cite{reagor_reaching_2013}, which is more pronounced in cavities with shorter wavelengths. We create our mode by intersecting several evanescent tubes as shown in Fig.~\ref{fig:Fig2}. Since an intersection of any two dissimilar bodies creates a pocket with a larger cross section than each of them separately, this yields a bound state below the cutoff of the tubes. Indeed, any arbitrarily weak defect in 1D has this property~\cite{chadan2003bound}, albeit with weaker localization. The number of evanescent tubes and their diameters determines the frequency of the resonance, while the locations of each intersection and angles between the tubes control the localization of the lowest mode and symmetries of higher order modes. Coupling into the cavity is set by the diameter and length of the input and output tubes attached to the WR10 waveguides. These parameters fully specify the seamless cavity, leaving the internal quality factor as the sole unknown. 

The flexibility of our design is demonstrated in Fig.~\ref{fig:Fig2}: using a single drill bit with a diameter below the cylindrical waveguide evanescent cutoff length $\frac{1.841\lambda }{2\pi}$, we are able to realize myriad resonator geometries with different spectra, depending on design requirements. The intersection of even two evanescent waveguides creates a well-localized mode at the ``elbow'' as shown in Fig.~\ref{fig:Fig2}a. Similarly, by intersecting more tubes in a ``tee'', Fig.~\ref{fig:Fig2}b, or ``star'' configuration, Fig.~\ref{fig:Fig2}c, we trap light at lower frequencies for the same diameter of the tube. For diameter $d=1.6mm$, a small number of tubes make conveniently sized high-Q resonators at mm-wave frequencies. This approach could be applied to coupled cavities and lattices of resonators, both in 2D and 3D, as shown in Fig.~\ref{fig:Fig2}e. Finally, many intersecting tubes can create lower frequency microwave cavities as shown in Fig.~\ref{fig:Fig2}d., where one can approach a large, seamless cylindrical or rectangular volume by drilling out the cavity tube by tube. For the purposes of hybrid experiments, evanescent tubes have the advantage of providing optical access for optical Fabry-P\'erot cavities, lasers and atomic beams as shown in Fig.~\ref{fig:Fig1}a.  All of these benefits make our design not only a powerful tool for hybrid Rydberg systems, but also a flexible tool for other circuit and cavity QED systems. 

The manufacturing of the device only involves drilling appropriate holes in high purity Niobium(Nb) stock. To reduce surface losses, the machined cavity is cleaned in solvents and chemically etched in a BCP bath of $2 H_{3}PO_{4}:HNO_{3}:HF$ for 20 minutes at room temperature~\cite{tian_surface_2006}. After rinsing and drying, we immediately mount the cavity inside a Helium-4 adsorption cryostat to avoid surface degradation due to oxidation. The system is cooled down to 1K, which is well below $T_c = 9.2$K of Nb, where the cavity is deep in the superconducting regime.

The 100GHz measurement setup is shown in Fig.~\ref{fig:Fig1}c. It is analogous to traditional microwave measurement chains, but with all components shifted to the 100GHz band. This includes: a Faraday isolator \textit{Quinstar QIF-W}, cryogenic amplifier \textit{LNF-LNC65-115WB} and directional coupler \textit{QJR-W-40-S}. For accessing the mm-wave band we use a multiplier \textit{VNAX-WR10} for up-conversion and multiplier + mixer \textit{VDI MixAMC 296} for down-conversion. We characterize our cavities using $S_{11}$ reflection to fit both magnitude and phase and thereby extract quality factors and resonance frequencieTo test the capabilities of our design, we have made and characterized several mm-wave cavities shown in Fig.~\ref{fig:Fig3}a. By varying the length and diameter of the in-coupling port as well as number of tubes at the intersection point, we are able to create resonances of varying frequency and coupling Q (Fig.~\ref{fig:Fig3}b). We consistently measured internal quality factors in the tens of millions. All of these cavities have mode volumes below $0.2\lambda^{3}$, which allows an tight confinement of mm-wave photons for tens of microseconds at 1K using evanescence of the tubes alone.

For the hybrid mm-wave cavity crossed with an optical Fabry-P\'{e}rot cavity for our experiments with Rydberg atoms in Fig.~\ref{fig:Fig1}a, we created the mm-wave mode volume by intersecting three tubes, each with diameter $1.5$mm. We chose the precise dimensions of our resonator to match the $35p$ to $36s$ transition frequency of the $^{85}$Rb (See App.~\ref{hybrid}) and to avoid clipping of the laser and atomic beams transiting the mode in the experiment. We measured an internal quality factor of $3 \times 10^7$ at $98.2$ GHz, which corresponds to $50 \mu s$ photon lifetime. To identify the loss channels limiting the lifetime of photons, we study how the internal quality factor depends on the power and temperature as shown in Fig.~\ref{fig:Fig3}c,d.

The internal quality factor of the resonator as a function of number of photons inside, shown in Fig.~\ref{fig:Fig3}c, reveals no power dependence from single photons to $\sim$10000 photons, indicating that our low-power Q is not limited by two-level system (TLS) absorbers, as is common in 2D resonators~\cite{khalil_loss_2011, martinis_decoherence_2005}, or that they remain unsaturated up to powers where other nonlinearities set in. We do observe a degradation in the quality factor and a nonlinear frequency response at high powers, suggesting the appearance of hot spots and meta-stable states on the surface of the etched Nb (see App.~\ref{NonlinearResponse}). As our primary interest lies in the performance of the cavity at single photon powers, we did not investigate this further.

Another common loss mechanism is surface resistivity due to thermal quasi-particles in the superconductor. Because this mechanism is known to be temperature dependent~\cite{gurevich_theory_2017}, we are able to rule it out by examining the behaviour of the resonator Q with temperature. In Fig.~\ref{fig:Fig3}d, we plot the temperature dependence of the fitted internal quality factor and fractional frequency change of the resonance for $\sim1$ and $9,000$ photons inside the cavity. The data is obtained by gradually heating and thermalizing the cavity using a power resistor. The internal quality factor as a function of temperature follows a Mattis-Bardeen law down to 2.3 K, where temperature-independent losses begin to dominate resulting in a deviation from the exponential trend. This indicates that we are not limited by thermal processes at the lowest temperatures. 

In addition to these mechanisms, other potential loss channels include magnetic flux pinning~\cite{romanenko_ultra-high_2014} and photon leakage at the coupling boundary. The performance of the cavity could be further improved by adding magnetic field shielding~\cite{ono_magnetic_1999} and sealing the rectangular to circular waveguide transition at the coupling port of the cavity to avoid leakage.

\begin{figure}[t!]
\centering
\includegraphics[width=1\linewidth]{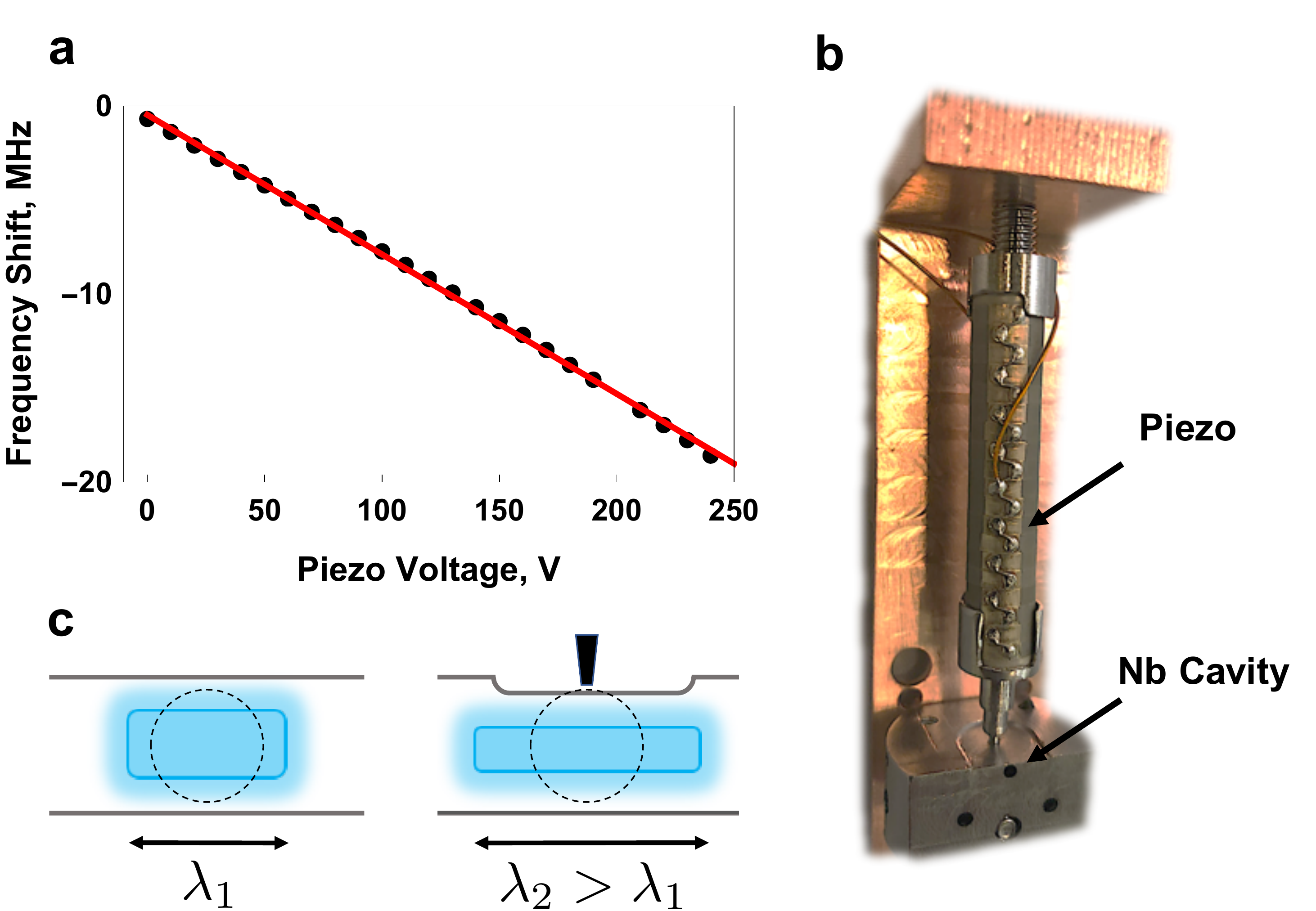}
\caption{Cryogenic frequency tuning of a high-Q mm-wave cavity.  
\textbf{a.} Frequency shift of the cavity resonance as a function of piezo voltage. At 1K, the linear fit shows $\sim 0.1$ MHz/V tunability, with maximum frequency shift of $\sim 18$MHz. At room temperature, the increased piezo throw enables cavity tuning by $\sim$GHz. \textbf{b.} Photograph of the piezoelectric actuator system attached to a test cavity. \textbf{c.} The low-temperature (1K) tuning is accomplished by displacing a pre-thinned wall of the cavity by  1-2 $\mu$m  using a piezo stack actuator. This pushes the mode farther out into the evanescent waveguides, effectively, increasing the wavelength and decreasing the frequency of the lowest mode.}
\label{fig:Fig4}
\end{figure}

To precisely match the frequency of the cavity to an atomic transition, the cavity must be tunable. To achieve this, we thin one side of the cavity and pre-load a \textit{Noliac} piezo electric stack actuator as shown in Fig.~\ref{fig:Fig4}b and c, to stress the cavity and thereby change its volume. Such stress-tuning provides several GHz of tunability at room-temperature, or $\sim 18$ MHz at 1K, as shown in Fig.~\ref{fig:Fig4}a. We note that while tunability is powerful, it has drawbacks: we observe that the thinning required to make the cavity tunable results in mechanical coupling to the pulse tube cryocooler vibrations, leading to fluctuations of the cavity frequency by many linewidths.

For hybrid experiments with Rydberg atoms, the seamless mm-wave cavity enabled immediate integration with an optical Fabry-P\'erot cavity with a waist $w=80\mu$m at 780nm and measured optical finesse $F = 10000$. For quantum interconversion between optical and mm-wave bands, large collective cooperativities on both optical and mm-wave transitions are desirable~\cite{CoveyMicrowaveConversion2019}; large single-particle cooperativity on the mm-wave transition will be an enabling technology for quantum-nonlinear optics~\cite{nikoghosyan2010photon}. We compute single-particle cooperativities of $C_{opt}=\frac{24F}{\pi}\frac{1}{(kw)^2}=0.2$ and $C_{mm}=22000$ (see App.~\ref{hybrid}), which satisfy these constraints for a moderate-density atomic ensemble~\cite{ningyuan2016observation}.

In conclusion, we have designed and machined a tunable mm-wave cavity with a measured internal Q factor of $3 \times 10^7$. The seamless geometry tightly confines photons at the intersection of evanescent waveguides in the $0.14{\lambda}^3$ mode volume, while allowing multiple lines-of-sight for optical access, injection of cold atoms and integration with an optical cavity for quantum interconversion experiments~\cite{gard_microwave--optical_2017,CoveyMicrowaveConversion2019}.

Our cavity Q's at 1K are comparable to those of state-of-the-art 3D microwave ($\sim 10$GHz) cavities at 20mK that are conventionally integrated with transmon qubits~\cite{reagor_reaching_2013, paik_observation_2011}. We anticipate the possibility of interfacing our high-Q, high-frequency resonators with such $10$GHz transmons via 100GHz nonlinear devices~\cite{anferov_millimeter-wave_2019, lotkhov_dc_2019} employed as mixers. 

For cavity and circuit QED, the tight confinement and long coherence of our device allows access to a strong coupling regime between a single photon and an emitter, which is difficult to achieve in many other platforms (see App.~\ref{Comparison}). The large cooperativity between a Rydberg atom and a mm-wave photon suggests studies of quantum manybody physics, photon-number-dependent group delay~\cite{nikoghosyan2010photon} and vacuum-induced squeezing~\cite{hu2017vacuum} on the mm-wave transition, while the effects can also be read out optically through the Fabry-P\'erot cavity. In sum, this work heralds a new era of optical/mm-wave quantum science.

%\end{itemize}
\begin{acknowledgments}
The authors would like to thank Andrew Oriani for help with the cryogenic setup. This work was primarily supported by the University of Chicago Materials Research Science and Engineering Center, which is funded by National Science Foundation under award number DMR-1420709. This work was supported by ARO grant W911NF-15-1-0397; D.S. acknowledges support from the David and Lucile Packard Foundation.

\end{acknowledgments}

\appendix

\section{\label{NonlinearResponse}Nonlinear frequency response of the cavity at high powers}

\begin{figure}[t!]
\includegraphics[width=1\linewidth]{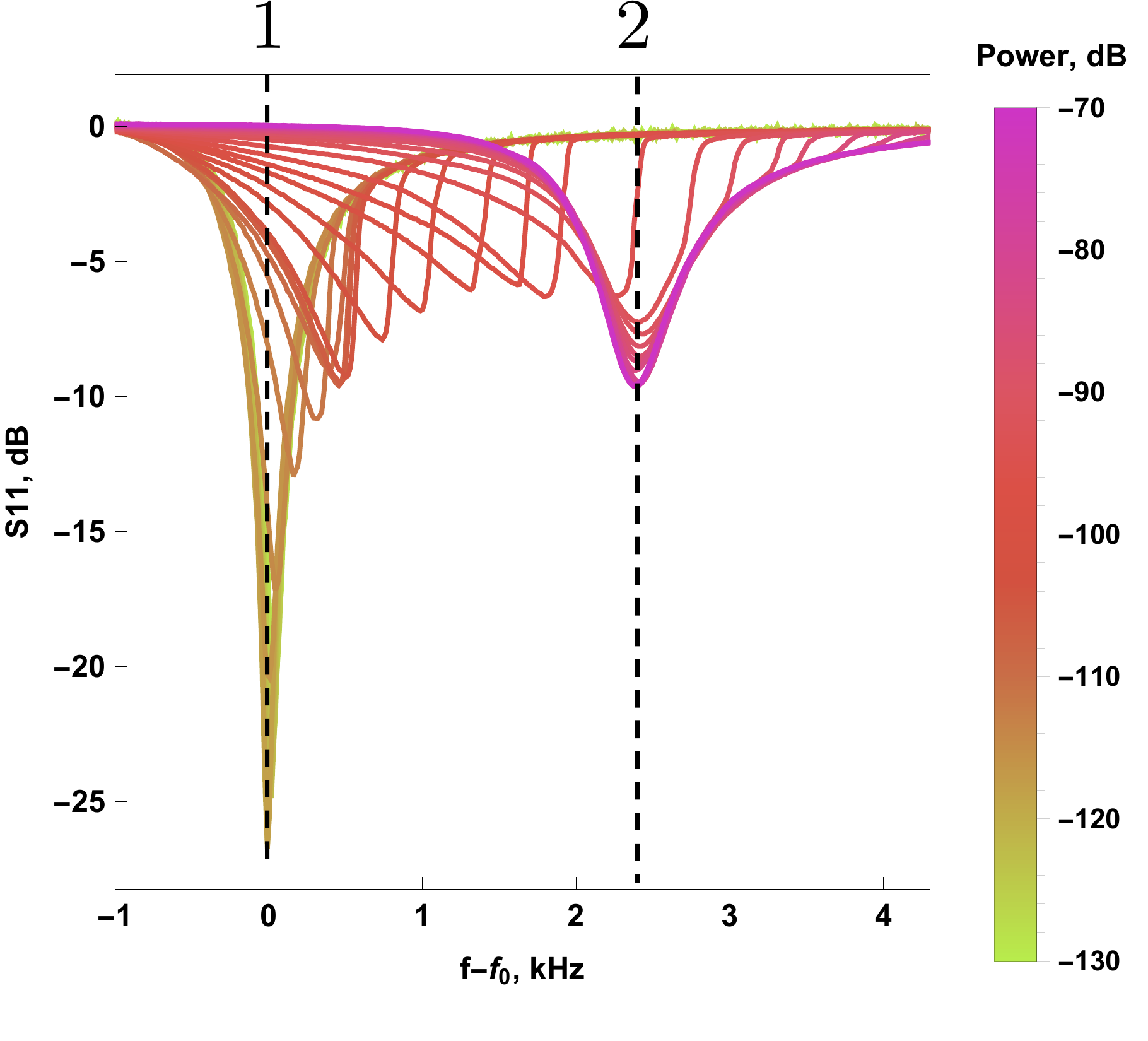}
\caption{The frequency response of the mm-wave cavity at different powers. There are two clear meta-stable states with linear frequency responses: 1 - at low powers with $f_{1} = 98.218508$ GHz and $Q_{1} = 3 \times 10^7$, and 2 - at high powers with $f_{2} = f_{1}+24$kHz and $Q_{2} = 1.3 \times 10^7$. For intermediate powers, the resonators exhibits Duffing-like behavior, suggesting nonlinearity caused by surface heating.}
\label{fig:FigS1}
\end{figure}

At high photon numbers, the frequency response of the cavity has a strong nonlinear character as shown in Fig.~\ref{fig:FigS1}. The nonlinear behavior appears at maximum simulated current densities of 100 A/cm$^{2}$, which is much smaller than the Nb critical current density of 150 MA/cm$^2$. This effect has been observed previously both in 2D~\cite{abdo_full_2013, oripov_high_2019, kurter_microscopic_2011} and in 3D~\cite{c._carvalho_piezoelectric_2016} superconducting resonators, and it is commonly attributed to non-linearity caused by hotspots or ``Josephon junction-like'' weak-links on the granular surface of Nb. The defects on the surface of granular superconductor have both superconducting and normal states, which drive the device into multiple metastable states: one with superconducting defects and one with normal metal defects at lower Q and different frequencies~\cite{abdo_full_2013, oripov_high_2019}. The magnetic flux penetration into the surface also plays an important role in this mechanism, and can enhance this effect making it more dramatic at high powers. Since our cavity is not shielded from stray B fields, we expect enhancement of the nonlinear behavior.

In Fig.~\ref{fig:FigS1}, at low intra-cavity powers around $P=-130$dBm or photon occupation of $n=10^4$ the cavity exhibits a linear response with a Lorentzian lineshape centered on $f_{1}=98.218508$ with the $Q_{1} = 3 \times 10^7$. As power increases, this state becomes metastable, and the frequency response exhibits a Duffing-like behavior, shifting up in frequency. At a power of $P\approx-80$dBm, the frequency shift saturates and the resonator falls into a second metastable state at $f_{2} = f_{1}+24$kHz with $Q_{2}= 1.3 \times 10^7$, again exhibiting a Lorentzian lineshape. It is important to note that even though the Duffing effect is common, a nonlinear coefficient resulting in a positive frequency shift with power is unusual. 

An understanding of the mechanism behind the frequency shifts, nonlinear behaviour and Q degradation in different power regimes is an ongoing topic of both experimental and theoretical research. The observed response at high powers is not prohibitive for our experiments in the quantum regime, since we will work at very low photon occupation number. If high power operation were required, shielding of the magnetic fields could inhibit flux pinning and other magnetic field effects contributing to loss. %Furthermore, increasing the number of tubes and decreasing the etch time would reduce the concentration of surface currents on the edges of the cavity mode, resulting in higher critical power for hot spots and the meta-stable behavior to appear.    

\section{\label{hybrid} Calculation of hybrid system cooperativities}

\begin{figure}[t!]
\includegraphics[width=0.85\linewidth]{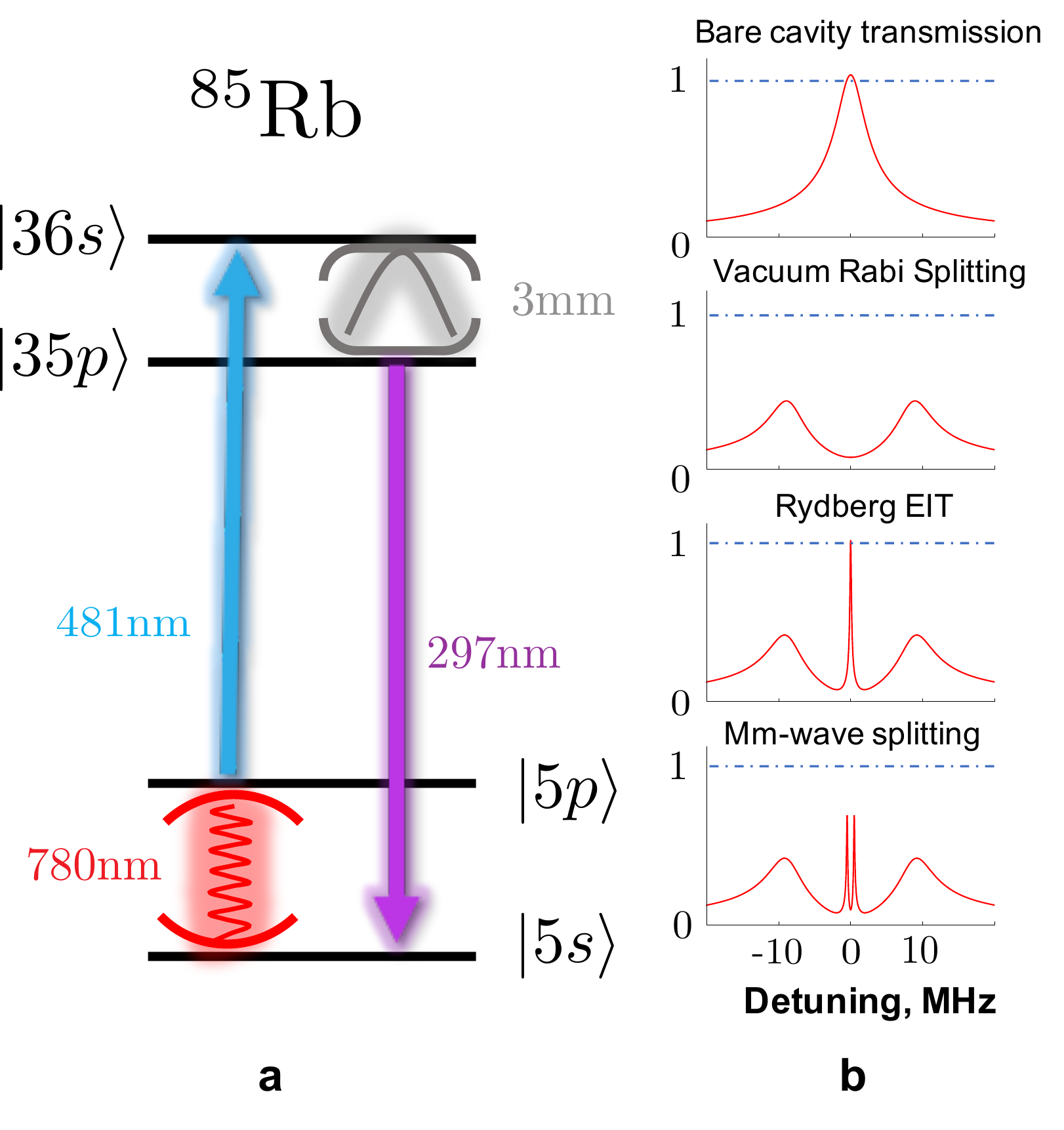}
\caption{The proposed hybrid cavity QED experiment with Rydberg atoms: \textbf{a} The energy levels of the $^{85}$Rb atoms involved in the hybrid experiments for interacting and coherently inter-converting mm-wave and optical photons. A blue laser is used to couple single mm-wave and optical photons, whereas an additional UV laser provides a way to bidirectionally convert between two bands. \textbf{b} The simulation of the optical cavity transmission in the hybrid experiment to observe the nonlinear interaction between single optical and mm-wave photons (top to bottom): starting with a single Lorentian peak corresponding to bare optical cavity transmission, then vacuum Rabi splitting of the cavity transmission due to presence of the atomic cloud, then EIT while the high power blue beam is on inside of the cavity mode, and, finally, mm-wave splitting due to presence of mm-wave photons inside of the resonator.}
\label{fig:FigS2}
\end{figure}

In this section we perform a rough estimate of the resonant coupling parameters for both the microwave and optical transitions of Rydberg atoms trapped at the center of the optical and mm-wave mode for the cavity shown in Fig.~\ref{fig:Fig1}. Fig.~\ref{fig:FigS2}a. shows the level diagram of four energy states of $^{85}$Rb we propose to utilize for both entangling and inter-converting single optical and mm-wave photons. The $780$nm optical photon in the Fabry-P\'erot cavity is resonant with ground $\ket{5S_{1/2}}$ to first excited state $\ket{5P_{3/2}}$ transition. The coupling strength and cooperativity between one Rb atom and a single optical photon are shown in Eq.~\ref{appa} and ~\ref{appb}:

\begin{eqnarray}
 \frac{g_{opt}}{2 \pi} = \frac{d_{opt}\cdot E_{opt}}{2 \pi \hbar} \approx 600 \text{ kHz}, \label{appa}
\\
C_{opt}= \frac{24F}{\pi}\frac{1}{(kw_0)^2}=0.2 \label{appb}
\end{eqnarray}

where $d_{opt}$ is the dipole moment $\bra{5s}er\ket{5p}$, $E_{opt}$ is the electric field strength of one optical photon at the location of the electron, $F$ is the finesse of the optical Fabry-P\'erot cavity, $k$ is the wavevector and $w_0$ is the waist of the optical cavity. The single atom interaction can be boosted by $\sqrt{N_{atoms}}$, due to coherent interaction between the cloud of $N_{atoms}$ cold atoms and a single photon~\cite{ningyuan2016observation}. 

On the mm-wave transition, the cooperativity between a Rydberg atom and a single mm-wave photon of the superconducting cavity is much higher due to strong confinement of 100GHz field in the cavity, which is shown in Eq.~\ref{appc} and ~\ref{appd}.

\begin{eqnarray}
 \frac{g_{mm}}{2 \pi} = \frac{d_{mm}\cdot E_{mm}}{2 \pi \hbar} = 460 \text{ kHz}, \label{appc}
\\
C_{mm}= \frac{4 g_{mm}^2}{\Gamma \kappa}=22000 \label{appd}
\end{eqnarray}
where $\Gamma 
$ here is the linewidth of the Rydberg energy state $\ket{36S}$ and $\kappa = \frac{f_0}{Q}$ is the linewidth of the mm-wave cavity. The high strength of the interaction is the result of the large Rydberg dipole moment $d_{mm} = \bra{35P}er\ket{36S}$ and a tight confinement of the mm-wave photon that our device provides.

The hybrid mm-wave/optical cavity in Fig.~\ref{fig:Fig1} allows for a cloud of cold atoms trapped in an optical lattice to enter through one of the tubes into, simultaneously, the center of the mm-wave cavity mode and the waist of Fabry-P\'erot cavity. Here, the atoms can interact efficiently with both mm-wave and optical photons. To facilitate this interaction, we use Electromagnetically-Induced Transparency (EIT) with a blue laser at $481$nm, which couples the $\ket{5P}$ and $\ket{36S}$ states. As shown in Fig~\ref{fig:FigS2}b., by weakly probing the optical cavity, we anticipate a vacuum Rabi splitting~\cite{zhu1990vacuum} of the optical transition due to presence of the cold atomic cloud, cavity EIT~\cite{zhang2008slow} in the presence of the blue light and finally, a splitting of the EIT peaks proportional to the square root of number of photons in the mm-wave cavity, akin to proposed photon-number dependent group velocity experiments in free space~\cite{nikoghosyan2010photon}. The strong coupling between single optical and mm-wave photons through interactions with atoms can open doors for entanglement and manipulation of mm-wave photons using optical light and vice versa. For interconversion of mm-wave and optical photons, we need a UV light at $297$nm between $\ket{35s}$ and $\ket{5s}$ states to allow coherent and bidirectional conversion and quantum information transfer.

\section{\label{Comparison}Comparison with other resonators}

Table.~\ref{tab:table1} contains a summary of different resonators types employed in cavity and circuit QED experiments, in both  optical and microwave regimes. To evaluate the performance of our device for quantum systems, we compared the mode volume and finesse of the resonators as parameters which are important for reaching strong-coupling regime. In cases of sub-wavelength microwave cavities, Q and finesse are the same. 

\begin{table*}[t]
\caption{\label{tab:table1} Common resonators in cavity and circuit QED systems}
\begin{ruledtabular}
\begin{tabular}{lccc}
 System &V, $\lambda^3$&f, GHz&Finesse\\ \hline
 2D superconducting resonators~\cite{leduc2010titanium}& $10^{-6}$ & 1.53 & $10^6 - 3\times 10^7$ \\
 3D superconducting resonators for qubits~\cite{reagor_reaching_2013}& 0.1 & 11 & $7\times 10^8$\\
 Our mm-wave cavity& 0.14 & 98.2 & $3\times 10^7$\\
 Accelerator Cavity~\cite{romanenko_understanding_2017}&1& 1-10& $10^{11}$\\
 MM-wave Fabry-P\'erot Cavity~\cite{kuhr_ultrahigh_2007}&260 & 51 & $4.6\times 10^9$\\
 Optical Fabry-P\'erot microcavity~\cite{vahala_optical_2003}& 6000 & $4.6\times 10^5$ & $10^5$\\
 Microsphere cavity~\cite{vahala_optical_2003,buck_optimal_2003}& 9000 & $3 \times 10^5$ & $10^6$\\
\end{tabular}
\end{ruledtabular}
\end{table*}

\nocite{*}
\bibliographystyle{naturemag}
\bibliography{References.bib}
\onecolumngrid
\newpage

\end{document}